# Superlubric Nanogenerators with Superb Performances

Xuanyu Huang[1], Li Lin[1], Quanshui Zheng[1,2*]

[1] Department of Engineering Mechanics and Center for Nano and Micro Mechanics, Tsinghua University, Beijing 100084, China

[2] State Key Laboratory of Tribology and Applied Mechanics Lab, Tsinghua University, Beijing 100084, China,

*E-mail: zhengqs@tsinghua.edu.cn

**Nanogenerators promise self-powered sensors and devices for extensive applications in internet of things, sensor networks, big data, personal healthcare systems, artificial intelligence, et al[1-5]. However, low electric current densities and short product lifespans have blocked nanogenerators' applications[3,6]. Here we** show that structural superlubricity, a state of nearly zero friction and wear between two contacted solid surfaces[7-11], provides a revolutionary solution to the above challenge. We investigate three types of superlubric nanogenerators (SLNGs), namely the capacitor-based, triboelectric, and electret-based SLNGs, and systematically analyze the influences of material and structural parameters to these SLNGs' performances. We demonstrate that SLNGs can achieve not only enduring lifespans, but also superb performances – three orders of magnitude in current densities and output powers higher than those of conventional nanogenerators. Furthermore, SLNGs can be driven by very weak external loads (down to ~1 $\mu$N) in very low frequencies (down to ~1 $\mu$Hz), and are thus capable to harvest electric energies from an extremely board spectrum of environments and biosystems. Among the three types of SLNGs, the capacitor-based is synthetically most competitive in the senses of performance, fabrication and maintaining. <span style="color:green">These results can guide designs and accelerate fabrications of SLNGs toward real applications.</span>

With the rapid developments of nanotechnology and microfabrication technology, ceaselessly miniaturized sensors and devices are emerging in vast numbers of applications in internet of things, sensor networks, big data, personal health systems, artificial intelligence, et al[1-5]. Until now, these sensors and devices have been mostly powered by batteries and external chargers, which limits their applications [12,13] particularly in needs for independent, sustainable, maintain-free operations of implantable biosensors, remote and mobile environmental sensors, nano/micro-scale robots or other electromechanical systems, portable/wearable person electronics, et al [13-16]. Because nanogenerators promise sensors and devices self-powered from environments and biosystems, they have been attracting rapidly increasing research interests since the first nanogenerator invented in 2006[1]. In particular, triboelectric nanogenerators (TENGs) have attracted the most intensively studies so far among all existing nanogenerators. A fundamental reason is that TENGs are based on a very common phenomenon – the friction electrification that is valid for most material pairs[17]. However, low electric current densities and short product lifespans of current nanogenerators have blocked real applications of nanogenerators explored so far[3,6,18]. This challenge was fundamentally caused by friction and wear[19,20].

Structural superlubricity (SSL) is a state of nearly zero friction and wear between two solid surfaces[8]. Since the first realization of microscale SSL in 2012 [7], SSL technology has been quickly growing with observations of SSL phenomena in high sliding speeds (up to 25 m/s)[21] and in a number of heterogeneous pairs of hBN[9], $MoS_2$[22], DLC[10], Au[11], and so on. These advances make superlubric nanogenerators (SLNGs) feasible. In this Letter we demonstrate that SSL technology[7-11,22] provides a revolutionary approach to invent nanogenerators with superb performances and enduring lifespans.

As illustrated in Figures 1(a) and 1(b), we first propose an elementary capacitor-based SLNG that is consist of three rectangular electrodes of same sizes (length $L$ and width $W$) separated by two dielectric films (DEF1 and DEF2) of thicknesses $d_1$ and $d_2$, and relative permittivity $\varepsilon_{r1}$ and $\varepsilon_{r2}$. The upper electrode which plays as the SLIDER is contacting with DEF1 in an SSL state which is the top part of the STATOR. In practice, for instance, a (conductive) graphite flake and a (dielectric) hexagonal boron nitride (hBN) film can be joint into a robust SSL contact [9]. Silicon dioxide ($SiO_2$) which is an insulator and can be grown directly on silicon by thermal oxidation or chemical vapor deposition process can provide another choice of the DEF1.

We take two steps to make the above setup into a SLNG. At the first step as illustrated in Figure 1(a), we collect the middle electrode to the lower (or equivalently, the upper) one by a circuit, apply a voltage, $V_c$, and then turn off the circuit. This step leads to the storage of electric charges $\pm Q_c$ in the two electrodes with $Q_c = C_2 V_c = \sigma_c S$, where $\sigma_c = \varepsilon_0 V_c / d_{r2}$ is the capacitor charge density, $\varepsilon_0 = 8.8542 \times 10^{-12} \text{F/m}$ is the vacuum permittivity, $S = LW$ and $C_2 = \varepsilon_0 S/d_{r2}$ are the area and capacitance of the capacitor, and $d_{r2} = d_2/\varepsilon_{r2}$ denotes the relative thickness of DEF2. The main aim of this step is to store and then isolate a charge in the middle electrode.

At the second step, the upper and lower electrodes are connected by a conductive circuit with an electric resistance, $R$, as illustrated in Figure 1(b). When moving the Slider, the changing capacitor, $C_1(t) = \varepsilon_0 S(1-z)/d_{r1}$, will result in a transferred charge, $Q(t)$, and current, $I(t) = Q'(t)$. Therefore, this elementary SLNG is in principle an electrostatic generator based on two capacitors, one changing $C_1(t)$ and one fixed $C_2$.

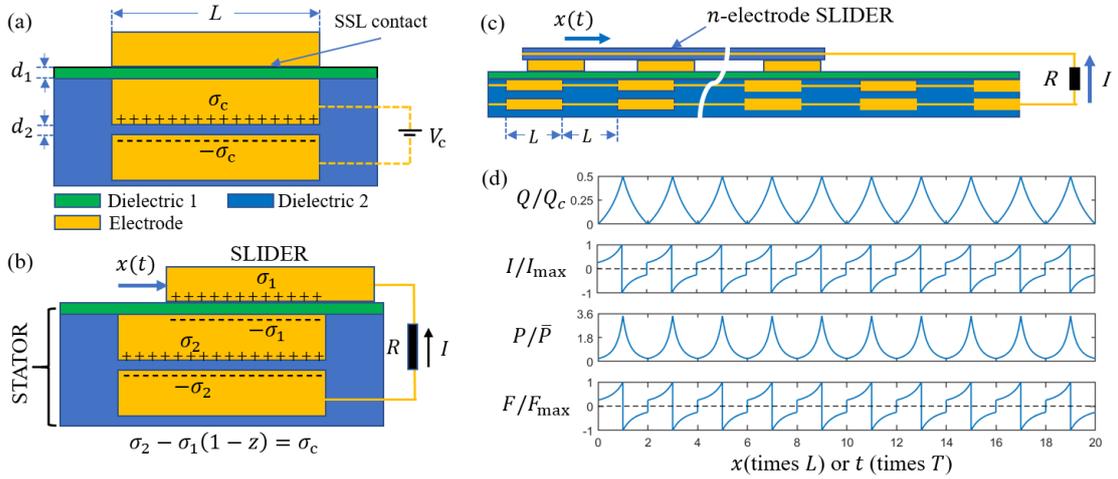

Figure 1. Basic capacitor-based SLNGs:
(a) The charging process and (b) the power generation process of an elementary capacitor-based SLNG where key variables and parameters are indicated, which is consist of a SLIDER and a STATOR contacted in a superlubric state. (c) The array form of capacitor-based SLNG. All the electrodes in each of the three layers are connected by a conductive wire to form an integrated larger equivalent electrode. (d) The relative transferred change $Q(t)/Q_c$, current $I(t)/I_{\max}$, output power $P(t)/\bar{P}$, and external force $F(t)/F_{\max}$ as functions of sliding distance or time when $\eta = 1$.

For simplicity we first consider a constant sliding speed, $v$. To realize this mode, the simplest manner is to extend the STATOR and SLIDER with periodically repeated and conductively connected three layers of electrodes as illustrated in Figure 1(c). As detailed in SI, the controlling differential equation and initial condition can be expressed as:

$$\frac{dy}{dz} + \alpha\left(1 + \left(1 + \frac{\eta}{1-z}\right)\left(y - \frac{1}{1+\eta}\right)\right) = 0, \qquad y(0) = 0, \tag{1}$$

where $y = Q(t)/Q_c$ denotes the relative transferred charge, $z = vt/L = t/T$ is the relative displacement or time, $T = L/v$ and $f = v/L$ are the across-electrode time and frequency, respectively,

$$\alpha = \frac{d_{r2}T}{\varepsilon_0 RS}, \tag{2}$$

and $\eta = d_{r1}/d_{r2}$ are two dimensionless parameters, and $d_{ri} = d_i/\varepsilon_{ri}$ denotes the relative thickness of DEF$i$ for $i = 1,2$.

The major purpose of this Letter is to explore the best possible performance of the capacitor-based SLNG. Thus, it is desired to have an explicit solution of Eq. (1). While the exact solution of Eq. (1) doesn't exist, by using theoretical analyses and numerical simulations as detailed in SI and illustrated in Figure 2(a-c), we luckily find that the solution $y$ as a function of $\alpha$ quickly approaches to a plateau (i.e. constant) as $\alpha$ is beyond roughly 100. For example, taking the sizes $L = W = 4$ μm, the silicon films as DEF1 and DEF2 ($\varepsilon_{r1} = \varepsilon_{r2} = 4.2$), the two equal thicknesses $d_1 = d_2 = 10$ nm, the sliding velocity $v = 1$ m/s, and the external series resistance $R = 100$kΩ, we have $\alpha = 672.27$. Since $\alpha \gg 1$ are valid for most practical cases mainly due to the extremely small $\varepsilon_0$, by taking $\alpha \to \infty$ in Eq. (1) we obtain the following explicit approximate solution:

$$y = \frac{1}{1+\eta} - \frac{1-z}{1+\eta-z}. \tag{3}$$

Further, based on Eq. (3) one can easily derive (as detailed in SI) the current, $I(t) = dQ(t)/dt$, the output power, $P(t) = RI(t)^2$, as well as the external force, $F(t)$, needed to drive the SLNG to generate electricity. These periodical solutions with respect to the setup in Figure 1(c) are plotted in Figure 1(d) when $\eta = 1$, in which $I_{\max}, \bar{P}$, and $F_{\max}$ denote the maximum current, the average output power, and the maximum required external force, respectively, that are summarized in the second column of the following Table 1. It is noted that both the transferred charge and output power are even functions and both the current and external force are odd functions over the one periodicity (e.g. $0 \leq z \leq 2$).

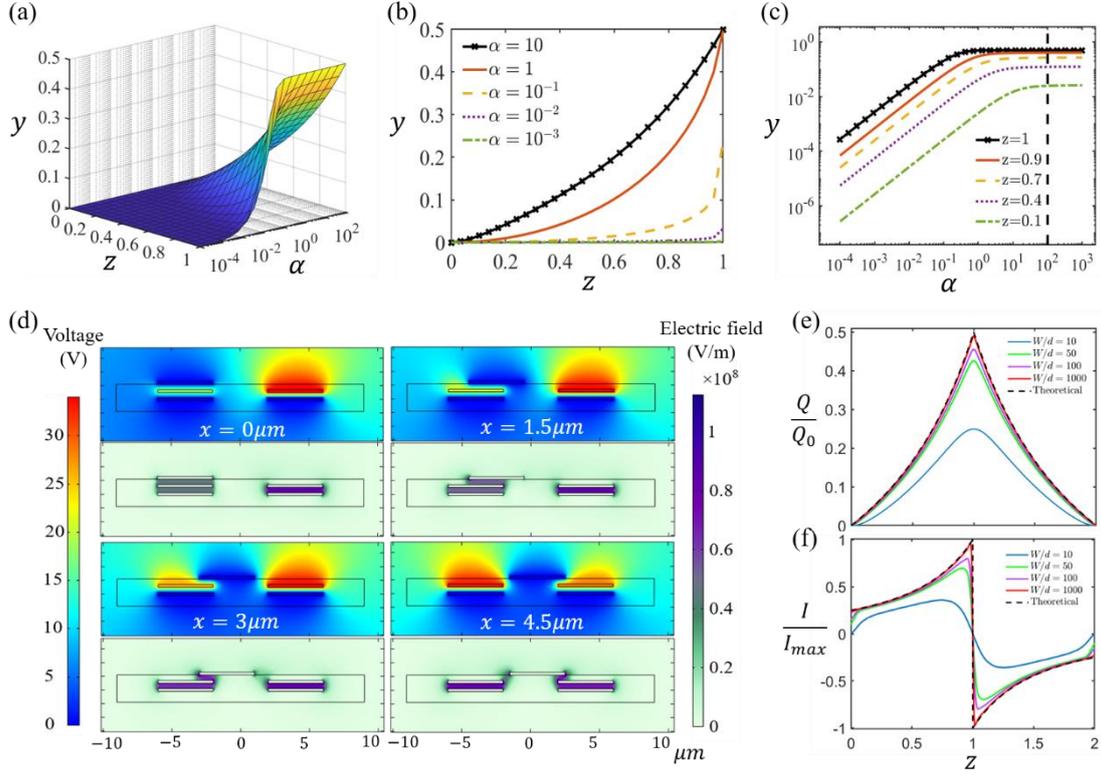

Figure 2. Numerical validation of the theoretical approximations of the capacitor-based SLNG with $\eta = 1$:
(a) The relief map of numerical solution $y(z)$ with relative sliding distance $z$ for different values of parameter $\alpha$. (b) The dependence of $y(z)$ upon the parameter $\alpha$ with respect to some specific $z$ values. (c) The dependence of $y(z)$ upon the parameter $z$ with respect to some specific $\alpha$ values. (d-f) The finite element simulation results: (d) The changes of potential field (uppers) and electric field (downs) with four displacements for a SLNG with $W/d = 10$. (e, f) The relative transferred charge and current as functions of the relative displacement $z = x/L$ with respect to four different width-thickness ratios, showing that the explicit approximations are excellent for $W/d > 50$.

Table 1. Performance comparisons among three types of nanogenerators.

| SLNG | Capacitor-based | Triboelectric | Electret-based |
|---|---|---|---|
| $Q_{max}$ | $\dfrac{1}{1+\eta}(\sigma_c S)$ | $\sigma_{TE} S$ | $\dfrac{1}{1+\eta}(\sigma_{EB} S)$ |
| $I_{max}$ | $\dfrac{1}{\eta}(f\sigma_c S)$ | $f\sigma_{TE} S$ | $\dfrac{1}{1+\eta}(f\sigma_{EB} S)$ |
| $\bar{P}$ | $\dfrac{1+3\eta+3\eta^2}{3(1+\eta)^3\eta}R(f\sigma_c S)^2$ | $R(f\sigma_{TE} S)^2$ | $\dfrac{1}{(1+\eta)^2}R(f\sigma_{EB} S)^2$ |
| $F_{max}$ | $\dfrac{1}{(1+\eta)\eta}\dfrac{d_0}{L}\dfrac{\sigma_c^2 S}{2\varepsilon_0}$ | $\dfrac{d_0}{L}\dfrac{\sigma_{TE}^2 S}{2\varepsilon_0}$ | $\dfrac{1}{(1+\eta)^2}\dfrac{d_0}{L}\dfrac{\sigma_{EB}^2 S}{2\varepsilon_0}$ |

To validate these explicit approximations, we carry out finite element simulations on the electric fields, potential fields, transferred charges, and currents as functions of the displacement $x$, as detailed in SI. With $L = W = 4\mu m$, $\varepsilon_{r1} = \varepsilon_{r2} = 4.2$, and $d_1 = d_2 = d$, Figure 2(d) shows the corresponding simulation results on the potential field (upper) and electric field (down) at four displacements for the capacitor-based SLNG with $W/d = 10$. Figures 2(e) and 2(f) show the simulation results (solid lines) on the relative transferred charge and current with respect to four different width-thickness ratios $W/d$. It is seen that the theoretical approximations (dashed line given by Eq. (3) and its derivative Eq.(S12)) are excellent for $W/d > 50$ because of reducing edge effects with increasing $W/d$, as detailed in SI.

A proof of concept of the above two-capacitor electrostatic generator is also performed by using the experimental setup shown in Figure 3(a). By controlling the rotational speed of the drive motor, we can easily realize an equivalent constant sliding speed $v$. The detailed experimental setup and the speed equivalence are given in SI, with experimental results of current $I(t)$ at some typical rotational speeds shown in Figure 3(b). The measured peak currents (dots with their error bars) are basically proportional to the rotatory speed that agree well with the theoretical prediction (red linear line).

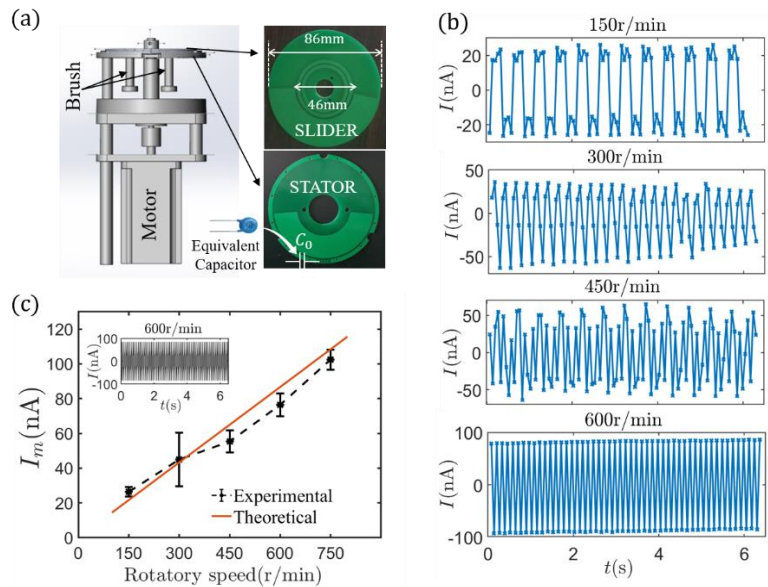

Figure 3. Proof of concept of the proposed two-capacitor generator: (a) Experimental setup. (b) Measured instant currents $I(t)$ at four typical angular speeds that correspond to four different sliding speeds. (c) Comparison with the theoretical predictions. The experimental error bars correspond to the current peaks at the first 20 rotating periods. The illustration is a theoretical calculation of the current waveform when the rotatory speed is 600 r/min.

The above validations accredit us to use the explicit results in the second column of Table 1 to explore the best potentials of the capacitor-based SLNG, that can be mainly characterized by the following two advantages:

**(1) Maximum possible pre-charge**. Because both $I_{\max}$ and $\bar{P}$ increase with the pre-charge density $\sigma_c$, a natural question is how to make $\sigma_c$ the maximum possible. Mathematically, one could unlimitedly enlarge $\sigma_c = \varepsilon_0 V_c/d_{r2}$ via either increasing the applied $V_c$, or decreasing $d_{r2}$. But a too high $\sigma_c$ will either break down the electric field strength limit, $E_{b2}$, of DEF2 during the capacitor pre-charge process, or break down the electric field strength limit, $E_{b1}$, of DEF1 during the SLNG working process. To avoid the happening in either case, the maximal allowed pre-charge density is found to be $\varepsilon_0 \min\{\eta\varepsilon_{r1}E_{b1}, \varepsilon_{r2}E_{b2}\}$, as detailed in SI. There is an optimized value of $\eta$, namely $\eta_{opt}=\varepsilon_{r2}E_{b2}/\varepsilon_{r1}E_{b1}$, or equivalently,

$$E_{b1}d_1 = E_{b2}d_2, \qquad (4)$$

at which we not only achieve the maximal allowed pre-charge density as follows:

$$\sigma_{c,\max} = \varepsilon_0\varepsilon_{r2}E_{b2}, \qquad (5)$$

but also, the best performances in transferred charge, current, and output average power (see SI). For instance, if we especially choose DEF1 and DEF2 to be the same and $d_1 = d_2$, then by taking $\eta_{opt} = 1$ we can achieve the best possible performances $I_{\max} = f\sigma_{c,\max}S$ and $\bar{P} = \frac{7}{24}R(f\sigma_{c,\max}S)^2$.

The above feature is of particular importance because we can freely choose DEF2 from dielectric materials that would have the highest or nearly highest $\sigma_{c,\max}$ (e.g. SiO$_2$, Al$_2$O$_3$, TiO$_2$, SrTiO$_3$, BaTiO$_3$, *et al.* [23,24]) without worrying about the SSL contact between the slider and DEF1.

**(2) Great size effect**. Because both $I_{\max}$ and $\bar{P}$ increase with $f = v/L$, we can therefore achieve excellent performances even at a very slow sliding speed $v$ (e.g. micrometers per second that corresponds an extremely low environmentally vibration frequency $\sim 1\ \mu Hz$) by using microscale $L$. Meanwhile, through shrinking $d_0$ from microscale into nano-scale we can reduce the required external force $F_{\max}$ in orders of magnitude.

Reducing $L$ (and $W$) leads to on one hand a higher $f = v/L$ and consequently higher densities of current and output power as indicated in the second column of Table 1, and on the other hand lower total current and output power because the faster reducing area $S = LW$. To solve this contradiction, we can use a large number (*n*) of the same electrodes in the SLIDER that are all connected by a metal wire, as illustrated in Figure 1(c). This approach is practical because of the microfabrication technology. The corresponding results can be directly given by using $S_n = nS$ to replace $S$ instead of these presented in the second column of Table 1 for a single-electrode SLIDER.

Triboelectric nanogenerators (TENGs) have been the most intensively investigated nanogenerators. Hereinafter we propose an elementary triboelectric SLNG as illustrated in Figure 4(a) that is a superlubric counterpart of the elementary TENG in the contact-sliding mode. Similar to the analytical approximations for the elementary capacitor-based SLNG, those for the triboelectric SLNG operated at a constant sliding speed $v$ can be derived, as detailed in SI. For an easier comparison, the corresponding explicit results of $Q_{\max}, I_{\max}, \bar{P}, F_{\max}$ are listed in the third column of Table 1. Here, $\sigma_{TE}$ denotes the triboelectric pre-charge density, the parameters $d_0 = d_{r1} + d_{r2}$, $\eta = d_{r1}/d_{r2}$, and $f = v/L$ are formally defined as the same as before, and the $S = LW$ denotes the area of the single element.

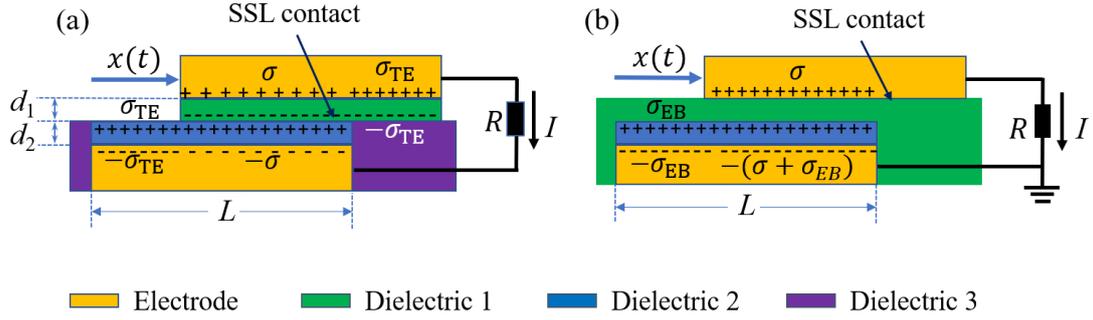

Figure 4 (a) The elementary triboelectric SLNG and (b) the elementary electret-based SLNG, where $\sigma_{TE}$ and $\sigma_{ET}$ denote the triboelectric and electret pre-charge densities, respectively.

The comparison of the second and third columns of Table 1 reveals that all the listed four performance indices for the capacitor-based and triboelectric SLNGs are formally similar, respectively. But practically, triboelectric SLNGs face physical and technical challenges in order to achieve a similar level of performances as those of capacitor-bases ones, as discussed below.

In physical aspect, the triboelectric pre-charge density $\sigma_{TE}$ is impossible to be higher than $\sigma_{c,max}$ because the latter as formulated in Eq. (5) is the theoretically maximum possible charge density before breaking down the DEF2. For instance, the values of $\sigma_{c,max}$ of the three dielectrics $Al_2O_3$, $TiO_2$, $BaTiO_3$ can be calculated as $2.41 \text{ mC/m}^2$, $17.7 \text{ mC/m}^2$, $49.6 \text{ mC/m}^2$,[23] in millimeter scale respectively. Furthermore, as a scale effect it is known that dielectrics breakdown strength $E_b$ increases with decreasing dielectric thickness in nanoscale[25]. Accordingly, the maximum value of $\sigma_{c,max}$ we calculate from the literature is around $708.3 \text{ mC/m}^2$ at $BaTiO_3$ [23]. In comparison, the maximum values of $\sigma_{TE}$ in air and vacuum we found from the literature are $0.25 mC/m^2$ and $1.00 \text{ mC/m}^2$, respectively, [26] that are three orders of magnitude lower than the maximum $\sigma_{c,max}$.

In technical aspect, it is hard to make a similar integrated structure for triboelectric SLNGs to that shown in Figure 1(b) for capacitor-based SLNGs. To preserve the triboelectric pre-charges in Dielectric 1, a third dielectric material (in purple color) has to be filled between the two Dielectric 2 flakes. However, it is a big technical challenge to make the surface composed of two different materials (Dielectrics 2 and 3) satisfying a fundamental SSL condition: atomically smooth without sharp steps of heights higher than one atomic layer[8]. In comparison, the top surface of the capacitor-based SNLG is a single dielectric film; while there are mature technologies to realize atomically smooth surfaces on single materials[27,28].

Electret materials have been often used to make electrostatic microgenerators[29-31]. Accordingly, we propose an elementary electret-based SLNG as illustrated in Figure 4(b), with an SSL contact between the upper electrode (SLIDER) and Dielectric 1 (the top surface of the STATOR). The corresponding results are presented in the fourth column of Table 1 and the detailed derivation is given in SI. Here, $\sigma_{EB}$ denotes the pre-charge density through high voltage to inject onto the top surface of the Dielectric 2. Electret-based SLNG has an advantage that we can make the maximum pre-charge density, $\sigma_{EB,max} = \varepsilon_0 \varepsilon_{r2} E_{b2}$, the same as given $\sigma_{c,max}$ before breaking down DEF2. Furthermore, as enlightened by the results shown in the fourth column of Table 1, we can achieve better $I_{max}$ and $\bar{P}$ through reducing the thickness of DEF1 (Optimize the parameter $\eta$).

However, the leakage is a common problem with the three types of nanogenerators mentioned above. As detailed in SI, the characteristic attenuation time of the charge is obtained as $\tau = \rho\varepsilon$, where $\rho$ and $\varepsilon$ are the resistivity and permittivity of dielectric film. For instance, use SiO$_2$ as dielectric, where $\rho \sim 10^{16} \Omega \cdot cm$ and $\varepsilon_r = 4.2$, so we can calculate the $\tau \approx 3.72 \times 10^5 s$, which is much larger than motion cycle time. Therefore, in the case of leakage, we can consider that other equations and derivations are completely unchanged, and only need to modify the capacitor charge density $\sigma_c$ to $\sigma_c e^{-\frac{t}{\tau}}$. In addition, the easier re-charging property of capacitor-based SLNG is a competitive advantage than the electret-based one.

All the above studied SLNGs have three common advantages. First, their open-circuit voltages can be reduced to a level of 1 V or even lower through shrinking the thicknesses of the dielectric films into nanoscale, in comparing with the often ultrahigh open-circuit voltages measured in contact-separation TENGs[32-34]. Nanogenerators with low open-circuit voltages are of particular importance for applications in biosystems. For example, by using a silicon oxide layer as DEF2 ($E_{b2} \approx 10^9$ V/m, $\varepsilon_{r2} \approx 4.2$), to achieve the maximal allowed $\sigma_{c,max}$ we need only to apply a low voltage $V_c = E_{b2}d_2 = 1$ V for the thickness $d_2 \approx 1$ nm. In comparison, if the thickness of the DEF2 $d_2 > 100$ μm that is typical in conventional nanogenerators[30,33], then one has to apply an extremely high voltage $V_c > 100,000$ V in order to achieve the maximal possible pre-charge $\sigma_{c,max}$.

Second, if the friction in a SSL contact would be negligible, then the theoretical analysis as detailed in SI shows that the mechanical-to-electric energy conversion efficiency can be higher than 97%. Even though in the ambient condition, the typical total friction force against sliding a 4-micrometer sized contact were measured to be in the range of ~ $1\mu N$[7-9,21]; and the friction from the edges contributes the majority and can be mostly removed through an annealing process[35]. In comparison, the electric or current-induced "friction" or resistance force against sliding includes is the electrostatic force $F$ formed by charge induction. This resistant force as represented in the Table 1 is in most practical cases significantly higher than $1\ \mu N$. For instance, if choosing the electrode sizes as $W = L = 4$μm, Al$_2$O$_3$ as Dielectric films 1 and 2 ($\varepsilon_{r1} = \varepsilon_{r2} = 5.7$), the thicknesses as $d_1 = d_2 = 100$ nm, the charging voltage as $V_c = 50V$, then the maximum resistant force can be calculated as $F_{max} = 10.09$ μN. Therefore, the efficiency of SLNG can be typically as high as ~90%.

Third, their required external force $F_{max}$ can be very low through shrinking $d_0$ from micro- into nano-scale, which therefore promises superb capability to harvest extremely weak and low-frequency mechanical energies from environments and biosystems. Nevertheless, for a comparison, the force exerted in the perpendicular mode is analyzed in SI for the TENG of contact-separation mode. The maximum external force to be overcome during the whole cycle is:

$$F_{max-c} = \frac{\sigma_{TE}^2 S}{2\varepsilon_0}. \tag{6}$$

It is $L/d_0$ times in the order of magnitude of those for the three studied SLNGs, and thus often corresponds to a large electrostatic force that limits applications. For instance, the required pull forces for contact-separation mode based on nanometer gaps exceeding $10^7 - 10^8$ N/m² [36].

Finally, let us consider the influence of randomicity of external loads that is common in environments. Instead of a constant-speed of the SLIDER shown in Figure 1(a), an overall oscillatory sliding characterized by $x(t) = \frac{L_{max}}{2}(1 - \cos\frac{2v_{max}t}{L_{max}})$ is considered, where $L_m$ denotes the maximum displacement distance of the SLIDER and $v_{max}$ the maximum instant speed. We find that the maximum transfer charge and the maximal or peak current has exactly the same form as in Table 1 while the average output power is about the half of its counterpart, where the frequency needs to be modified to $f_{max} = v_{max}/L$. Consequently, these results confirm that the capacitor-based SLNG work well in random environment.

To summarize, we propose and study three elementary types of SLNGs. They promise a great capability of harvesting weak, low-frequency, and random mechanical energies from environments and biosystems. Their performances can reach to at least three orders of magnitude higher than those of conventional nanogenerators where superlubricity doesn't apply. In particular, combining SSL and nanofabrication technologies, miniaturized SLNGs are promising, with extra high energy convert efficiency because of nearly zero friction. Therefore, commercialization of SLNGs can be expected to a reality in a near future because of superlubricity, which will then lead to their broad applications.

# References

1 Wang, Z. L. & Song, J. H. Piezoelectric nanogenerators based on zinc oxide nanowire arrays. *Science* **312**, 242-246 (2006).
2 Li, A., Zi, Y., Guo, H., Wang, Z. L. & Fernandez, F. M. Triboelectric nanogenerators for sensitive nano-coulomb molecular mass spectrometry. *Nature Nanotechnology* **12**, 481-487 (2017).
3 Liu, J. *et al.* Direct-current triboelectricity generation by a sliding Schottky nanocontact on MoS2 multilayers. *Nature Nanotechnology* **13**, 112-+, (2018).
4 Qin, Y., Wang, X. & Wang, Z. L. Microfibre-nanowire hybrid structure for energy scavenging. *Nature* **457**, 340-340 (2009).
5 Yang, R., Qin, Y., Dai, L. & Wang, Z. L. Power generation with laterally packaged piezoelectric fine wires. *Nature Nanotechnology* **4**, 34-39 (2009).

6   Wang, Z. L. Triboelectric nanogenerators as new energy technology for self-powered systems and as active mechanical and chemical sensors. *Acs Nano* **7**, 9533-9557 (2013).

7   Liu, Z. *et al.* Observation of microscale superlubricity in graphite. *Physical Review Letters* **108** (2012).

8   Hod, O., Meyer, E., Zheng, Q. & Urbakh, M. Structural superlubricity and ultralow friction across the length scales. *Nature* **563**, 485-492 (2018).

9   Song, Y. *et al.* Robust microscale superlubricity in graphite/hexagonal boron nitride layered heterojunctions. *Nature Materials* **17**, 894-+ (2018).

10  Berman, D., Deshmukh, S. A., Sankaranarayanan, S. K. R. S., Erdemir, A. & Sumant, A. V. Macroscale superlubricity enabled by graphene nanoscroll formation. *Science* **348** (2015).

11  Kawai, S. *et al.* Superlubricity of graphene nanoribbons on gold surfaces. *Science* **351**, 957-961 (2016).

12  Shao, H., Tsui, C.-Y. & Ki, W.-H. The design of a micro power management system for applications using photovoltaic cells with the maximum output power control. *Ieee Transactions on Very Large Scale Integration (Vlsi) Systems* **17**, 1138-1142 (2009).

13  Wang, Z. L. & Wu, W. Nanotechnology-enabled energy harvesting for self-powered micro-/nanosystems. *Angewandte Chemie-International Edition* **51**, 11700-11721 (2012).

14  Fan, F. R., Tang, W. & Wang, Z. L. Flexible nanogenerators for energy harvesting and self-powered electronics. *Advanced Materials* **28**, 4283-4305 (2016).

15  Mahmud, M. A. P., Huda, N., Farjana, S. H., Asadnia, M. & Lang, C. Recent advances in nanogenerator-driven self-powered implantable biomedical devices. *Adv. Energy Mater.* **8**, 25 (2018).

16  Proto, A., Penhaker, M., Conforto, S. & Schmid, M. Nanogenerators for human body energy harvesting. *Trends Biotechnol.* **35**, 610-624 (2017).

17  Wang, Z. L. & Wang, A. C. On the origin of contact-electrification. *Materials Today* (2019).

18  Lin, S., Lu, Y., Feng, S., Hao, Z. & Yan, Y. A high current density direct-current generator based on a moving van der Waals Schottky diode. *Advanced Materials* **31** (2019).

19  Lin, L. *et al.* Robust Triboelectric nanogenerator based on rolling electrification and electrostatic induction at an instantaneous energy conversion efficiency of similar to 55%. *Acs Nano* **9**, 922-930 (2015).

20  Tang, W. *et al.* Liquid-metal electrode for high-performance triboelectric nanogenerator at an instantaneous energy conversion efficiency of 70.6%. *Advanced Functional Materials* **25**, 3718-3725 (2015).

21  Yang, J. *et al.* Observation of high-speed microscale superlubricity in graphite. *Physical Review Letters* **110** (2013).

22  Wang, L. F. *et al.* Superlubricity of a graphene/MoS2 heterostructure: a combined experimental and DFT study. *Nanoscale* **9**, 10846-10853 (2017).

23  Neusel, C. & Schneider, G. A. Size-dependence of the dielectric breakdown strength from nano- to millimeter scale. *Journal of the Mechanics and Physics of Solids* **63**, 201-213 (2014).

24  Qi, T., Irwin, P. & Yang, C. Advanced dielectrics for capacitors. *Transactions of the Institute of Electrical Engineers of Japan, Part A* **126-A**, 1153-1159 (2006).

25  Chen, G., Zhao, J., Li, S. & Zhong, L. Origin of thickness dependent dc electrical breakdown in dielectrics. *Applied Physics Letters* **100** (2012).

26  Wang, J. *et al.* Achieving ultrahigh triboelectric charge density for efficient energy harvesting. *Nature Communications* **8** (2017).

27  Chen, L. *et al.* Nanomanufacturing of silicon surface with a single atomic layer precision via mechanochemical reactions. *Nature Communications* **9** (2018).

28  Zhang, X., He, Y., Li, R., Dong, H. & Hu, W. 2D mica crystal as electret in organic field-effect transistors for multistate memory. *Advanced Materials* **28**, 3755-3760 (2016).

29  Beeby, S. P., Tudor, M. J. & White, N. M. Energy harvesting vibration sources for microsystems applications. *Measurement Science and Technology* **17**, R175-R195 (2006).

30  Nguyen, C. C., Ranasinghe, D. C. & Al-Sarawi, S. F. Analytical modeling and optimization of electret-based microgenerators under sinusoidal excitations. *Microsystem Technologies-Micro-and Nanosystems-Information Storage and Processing Systems* **23**, 5855-5865 (2017).

31  Suzuki, Y. Recent progress in MEMS electret generator for energy harvesting. *IEEJ Trans. Electr. Electron. Eng.* **6**, 101-111 (2011).

32  Liu, J. *et al.* Sustained electron tunneling at unbiased metal-insulator-semiconductor triboelectric contacts. *Nano Energy* **48**, 320-326 (2018).

33  Niu, S. *et al.* Theoretical study of contact-mode triboelectric nanogenerators as an effective power source. *Energy & Environmental Science* **6**, 3576-3583 (2013).

34  Wang, Z. L., Chen, J. & Lin, L. Progress in triboelectric nanogenerators as a new energy technology and self-powered sensors. *Energy & Environmental Science* **8**, 2250-2282 (2015).

35  Wang, W. *et al.* Measurement of the cleavage energy of graphite. *Nature Communications* **6**, 7 (2015).

36  Kostsov, E. G. Ferroelectric-based electrostatic micromotors with nanometer gaps. *Ieee Transactions on Ultrasonics Ferroelectrics and Frequency Control* **53**, 2294-2298 (2006).